# High resolution ultra-sensitive trace gas detection by use of cavity-position-modulated sub-Doppler noise-immune cavity-enhanced optical heterodyne molecular spectrometry – Application to detection of acetylene in human breath

GANG ZHAO,[1,2,3] THOMAS HAUSMANINGER,[1] FLORIAN M. SCHMIDT,[4] WEIGUANG MA,[2,3] AND OVE AXNER[1,*]

[1]*Department of Physics, Umeå University, SE-901 87 Umeå, Sweden*
[2]*State Key Laboratory of Quantum Optics & Quantum Optics Devices, Institute of Laser Spectroscopy, Shanxi University, 030006 Taiyuan, China*
[3]*Collaborative Innovation Center of Extreme Optics, Shanxi University, 030006 Taiyuan, China*
[4]*Department of Applied Physics and Electronics, Umeå University, SE-901 87 Umeå, Sweden*
*\* ove.axner@umu.se*

**Abstract:** A sensitive high resolution sub-Doppler detecting spectrometer, based on noise-immune cavity-enhanced optical heterodyne molecular spectrometry (NICE-OHMS), for trace gas detection of species whose transitions have severe spectral overlap with abundant concomitant species is presented. It is designed around a NICE-OHMS instrumentation utilizing balanced detection that provides shot-noise limited Doppler-broadened detection. By synchronous dithering the positions of the two cavity mirrors, the effect of residual etalons between the cavity and other surface in the system could be reduced. An Allan deviation of the absorption coefficient of $2.2 \times 10^{-13}$ cm$^{-1}$ at 60 s, which, for the targeted transition in C$_2$H$_2$, corresponds to a 3σ detection sensitivity of 130 ppt, is demonstrated. It is shown that despite significant spectral interference from CO$_2$ at the targeted transition, which precludes Db detection of C$_2$H$_2$, acetylene could be detected in exhaled breath of healthy smokers.

## 1. Introduction

Laser absorption spectroscopy (LAS) is a powerful technique for trace gas detection that is often characterized by high sensitivity, selectivity and accuracy [1]. It has therefore been successfully applied to a number of fields, including environmental monitoring [2], industrial process control [3-5], defense and homeland security [6], and medical diagnostics [7, 8].

Although the most basic realization of LAS, direct absorption spectroscopy (DAS), has a number of advantages—it can be made robust, compact, and calibration-free—it is limited in terms of sensitivity due to low-frequency 1/*f* - type of noise. More sensitive realizations of LAS have therefore been developed. Various modulation techniques have been used to move the analytical information to higher frequencies. The most common ones are wavelength and frequency modulation (WMS and FMS) [9, 10]. An alternative means is to extend the interaction length between the light and analyte by use of a multipass cell [11] or a resonant cavity [12]. This has demonstrated effective absorption lengths of km in laboratory environments and led to the development of a variety of cavity enhanced detection techniques, e.g. cavity enhanced absorption spectroscopy (CEAS) [13], cavity ring down spectroscopy (CRDS) [14-16], and off-axis integrated cavity output spectroscopy (OA-ICOS) [17].

Despite the development of such techniques, the detection of gas molecules in trace concentrations is, in many types of applications, often hampered by overlapping spectral lines from concomitant species, so called spectral interferences. A necessary prerequisite for dealing with this is a high selectivity. Two possible means to separate transitions is to use multi-line fitting (to extract information about the analyte from measurement data originating from all species) or to detect the sample under low pressure conditions. The latter action leads to a decrease of the linewidth from the GHz range, which many transitions exhibits under atmospheric pressure conditions, to the Doppler limit, which, in the tenths of Torr region or below, typically is in the hundreds of MHz range [18, 19].

In many cases, in particular when the concomitant species are either present in high concentrations or have strong transitions in the same spectral region, these actions are still insufficient for accurate quantification of trace gases. To further improve on the spectral resolution, sub-Doppler (sD) spectroscopy can be used. This technique is built upon optical saturation of the target transition, e.g. by the use of counter propagating laser beams, so as to remove the influence of the inhomogeneous Doppler-broadening. By this, the linewidth of the transition can be narrowed to the MHz range [20, 21].

However, for molecular transitions in the near-IR (NIR) wavelength range, in which there are a plentitude of optical, electro-optical, and acousto-optical devices available that facilitates the realization of instrumentation working in this wavelength range, the dipole moments are normally low, often below the mD-region [22]. This implies that high laser powers, often in the tens of W region, paired with low pressure conditions, often in the sub-Torr pressure range, are needed to saturate a transition. Such high laser powers are difficult to achieve in practice, and the low sample pressures required preclude the measurement of low concentrations of analytes using single-pass configurations. On the other hand, the high intracavity power in resonant cavities, paired with the extraordinary long interaction lengths, open up for sD detection for weak molecular transitions in the NIR range (including those that possess small dipole moments). However, despite this potential, it has been found that the ability of the aforementioned cavity enhanced detection techniques to utilize sD spectroscopy for trace gas detection is often restricted because of technical reasons; CEAS is limited by the frequency-to-amplitude noise conversion [23], CRDS is restricted by the discrete nature of the frequencies of the cavity modes, while OA-ICOS is hampered by its reduced output intensity [24]. Despite a few such demonstrations [25-27], sD spectroscopy has therefore so far not been seen as a viable option for trace molecular species using any of these techniques.

However, in the end of the 1990's, based on a combination of CEAS and FMS, a laser-based technique with a unique set of properties, among them an astonishing detection sensitivity, referred to as noise-immune cavity-enhanced optical heterodyne molecular spectrometry (NICE-OHMS), was developed [28-30]. In this, the carrier of the modulated laser is locked to one of the longitudinal modes of the cavity [most often a Fabry-Perot (FP) cavity] by the Pound-Drever-Hall (PDH) technique [31], while the modulation frequency is locked to the cavity free spectral range (FSR) by the DeVoe-Brewer (DVB) technique [32]. By this, NICE-OHMS acquires both an extraordinary sensitivity and a unique immunity to frequency-to-amplitude noise conversion [30, 33, 34]. In fact, the technique was originally developed to address weak sD features of $C_2HD$ for frequency reference purposes [28-30].

Since its earliest realizations, the NICE-OHMS technique has mainly been developed around various types of tunable lasers. To expand its applicability to trace gas detection, it has predominately addressed Db transitions [33-55]; it has only occasionally been utilized for sD detection, then for other purposes [20, 56-60]. Among the latter realizations, a white noise limited detection sensitivity of $5 \times 10^{-11}$ cm$^{-1}$Hz$^{-1/2}$ has been reported [57]. Despite these demonstrations, and the fact that, primarily due to its MHz resolution, NICE-OHMS with sD mode of detection provides a large potential for trace gas detection [20], sD NICE-OHMS has up to now, to the authors' knowledge, not been applied to this field of applications.

A significant amount of development of the NICE-OHMS technique has been performed based on Erbium-doped fiber lasers (EDFL). Due to their narrow linewidth and intrinsic stable performance, a number of realizations of the NICE-OHMS technique, based on such lasers, with successively improved detection sensitivities for Db detection have been presented in a series of works [39, 40, 43, 46-49, 54, 55]. Most recently, based on balanced detection, a system for shot noise limited Db detection with an unprecedented detection sensitivity of $2.2 \times 10^{-14}$ cm$^{-1}$ (for an integration time of 200 s) was demonstrated [55]. Despite the latter achievement, the experimental system used in that work was found not to be shot noise limited for sD detection. The reason was attributed to the appearance of narrow structures in the background, caused by etalons generated between one of the mirrors of the cavity and a second surface in the system, with a periodicity similar to the width of the sD NICE-OHMS signal. This implies that no NICE-OHMS system has so far been able to utilize its full power for trace gas detection based on sD detection. This has hampered some of the most interesting types of applications. There is therefore a continuous strive to improve on this situation.

One means to reduce the influence of etalons, originally demonstrated for tunable diode laser absorption spectroscopy (TDLAS) [61], and more recently applied to cavity-enhanced spectroscopy of molecular ions [62], is to dither one of the components between which the etalon appears. Based on this principle, and as is presented in this paper, an alternative means to reduce the influence of etalons is to *simultaneously* move both mirrors of the FP cavity. By this, the length of any etalon generated between one of the cavity mirrors and any other surface in the system can be periodically modulated without noticeably altering the length of the cavity, i.e. without shifting any cavity mode. Since this method utilizes the piezoelectric transducers (PZTs) of the cavity, which are used to mount the cavity mirrors to the spacer and to control the frequencies of the cavity modes, no additional optical components need to be added to the system. The performance of this novel method has been scrutinized by measurements of background signal analyzed by the use of Allan-Werle plots.

It was then specifically investigated whether this system could be applicable for monitoring of $C_2H_2$ in breath. Laser-based absorption techniques have previously demonstrated to be powerful for detection of molecular species in human breath [7, 8, 63, 64]. Regarding acetylene, recent studies have shown that a single cigarette can release about 150 μg of acetylene to the environment [65], and thus much higher $C_2H_2$ concentrations are exhaled by smokers than non-smokers [66, 67]. Directly after smoking or acute exposure to air pollution, the breath $C_2H_2$ levels can reach hundreds of ppb, followed by a relatively fast elimination (exponential decay) from the body within about 3 hours [66, 67].

From the aforementioned breath acetylene studies, which were performed by the use of CRDS and OA-ICOS in the NIR region, it is evident that other breath constituents, predominantly the highly abundant species $H_2O$, $CO_2$, $CH_4$, and $NH_3$, often give rise to spectral interferences when $C_2H_2$ is detected. In fact, in some wavelength regions, e.g. that covered by the EDFL laser used in this work, no $C_2H_2$ transition has been found that does not suffer from significant interferences from at least one of the aforementioned species.

In this work, despite the presence of significant spectral interference from $CO_2$, which precludes or significantly deteriorates Db detection, sD NICE-OHMS was used to detect low concentrations of $C_2H_2$ in breath. In addition, since assessments of minor constituents in breath are of relevance only if their concentrations are related to the total amount of breath exhaled, which use to be assessed by monitoring the concentration of $CO_2$, a $C_2H_2$ transition was selected that was in close proximity to (i.e. on the wing of) a $CO_2$ signal. By this, the concentration of $C_2H_2$ could be assessed by sD spectroscopy, while the concentration of $CO_2$ could be monitored as the Db part of the signal on which the sD signal resides. The applicability of this approach is demonstrated by measuring the concentration of acetylene in the breath of two healthy human subjects after smoking.

## 2. Experimental setup

The experimental setup, which is shown in Fig. 1, is, to a large extent, based on that for balanced detection Db NICE-OHMS presented in Ref. [55]. In short, the system is constructed based on an EDFL that operates at around 1534 nm, addressing an acetylene transition at 6518.4858 cm$^{-1}$ with a line strength of $8.572 \times 10^{-21}$ cm$^{-1}$/(mol cm$^{-2}$) and a dipole moment of 8.0 mD. The output of the laser is passed through a fiber-coupled acoustic optic modulator (*f*-AOM), a fiber-coupled polarizer (*f*-POL), and a fiber-coupled electro-optic modulator (*f*-EOM), the latter with a proton exchanged waveguide to minimize the generation of RAM [68] before it is emitted into free space by the use of a fiber-coupled collimator (*f*-C). The frequency of the light is controlled by both a PZT inside the laser and a radio frequency (RF) controlled *f*-AOM (by the use of a tunable VCO). The laser light is phase modulated by sending two radio frequencies to the *f*-EOM (one at 20 MHz with a modulation index of 0.1 for the PDH locking, and one at 380 MHz with a modulation index of 1 for the FMS).

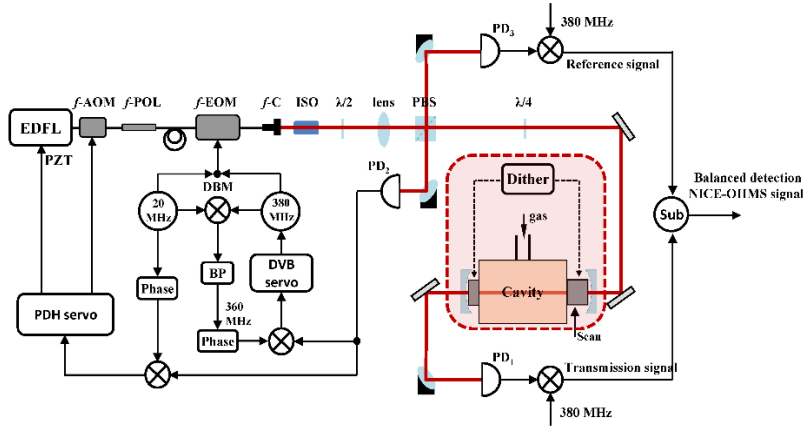

Fig. 1. Experimental setup. EDFL, erbium-doped fiber laser; *f*-AOM, fiber-coupled acousto-optic modulator; *f*-POL, fiber-coupled polarizer; *f*-EOM, fiber-coupled electro-optic modulator; *f*-C, fiber-coupled collimator; ISO, isolator; λ/2; half-wave plate; PBS, polarizing beam splitter; λ/4, quarter-wave plate; $PD_1$, $PD_2$, and $PD_3$, photodiodes; PS, power splitter; DBM, double balanced mixer; Sub, subtraction.

The light is then coupled into a FP cavity with a finesse of 55 000 by the use of a mode matching lens. The transmitted light is focused by a parabolic mirror onto a photodetector ($PD_1$). To yield the transmission NICE-OHMS signal, this detector signal is demodulated at the modulation frequency (380 MHz) by the use of a mixer (at a phase that provides a dispersive response). The cavity reflected light, which is deflected from the impinging light by the use of a polarizing beam splitter (PBS) cube, is focused onto a second photodetector ($PD_2$), whose signal is used for the PDH and DVB lock.

As was alluded to above, to reduce the amount of background signals generated in the optical system before the PBS, especially in the *f*-EOM, a balanced detection scheme was implemented [55]. A part of the incoming light, deflected by the PBS, was detected by a third photodetector ($PD_3$), whose signal was demodulated in the same way as the signal from the photodetector detecting the transmission NICE-OHMS signal, i.e. $PD_1$, to create a reference signal. The NICE-OHMS background signal could then be effectively reduced by a subtraction of the reference NICE-OHMS signal from the corresponding signal measured in transmission, with the former weighted by the ratio of laser powers in transmission and reflection.

However, even when balanced detection was used, it was found that there were some remaining structures in the background signal through which noise couples in. These were attributed to etalons produced after the PBS, either in the transmission or in the reference arm, between one of the cavity mirrors and another surface in the system. Although the magnitude of such etalons can be reduced by the use of etalon immune distances (EIDs) [47], they can still appear if the conditions for EID are not strictly fulfilled. To reduce the remaining etalons, the positions of the two cavity mirrors were modulated synchronously by imposing two 90 Hz sinusoidal signals with a fixed phase-relation to the two cavity PZTs. By this the entire cavity was dithered, while its length was not markedly altered. This implies that the cavity mirrors were dithered without discernibly changing the cavity modes.

However, it was found that, even after a careful adjustment of the amplitudes and the relative phase of the two 90 Hz dither signals, it was not possible to completely remove the influence of the dithering on the cavity length. This was attributed to nonlinear modulation responses of the PZTs, which were different for the two PZTs. To compensate for this, a weak second harmonic signal (i.e. at 180 Hz) was added to one of the PZTs. The phases and amplitudes of the various dither signals were then adjusted until there was neither any 90 Hz, nor any 180 Hz, signal in the feedback signals of the PDH locking servos, which indicated that the dithering did not noticeably affect the cavity length.

The amplitude of the dithering was chosen so that the shifts of the etalons were larger than their FSR. By this, an averaging of the signal could efficiently be used to reduce the effect of the remaining etalons in the background signal.

### 3. System evaluation – reduction of influence of background signals

To evaluate the performance of the system, Fig. 2 shows the measured drift of some NICE-OHMS background signals recorded at dispersion phase. The data represent the difference between two empty cavity measurements (each averaged over 90 scans, each produced by a triangle wave with a period of 1.4 s) taken at two different time instances, separated by 4 hours, henceforth referred to as the drift of the signal. The red curves in the panels (a) – (c) display the drift of the background signals for three different modes of detection, *viz.* when the NICE-OHMS signal: *i)* is solely taken as that from the transmission detector (thus without any dithering of the mirrors), representing "ordinary" NICE-OHMS; *ii)* originates from balanced detection without dithering, and *iii)* is assessed by balanced detection with dithering, respectively. The black curve in each panel represents the Db NICE-OHMS background signal obtained by fitting.

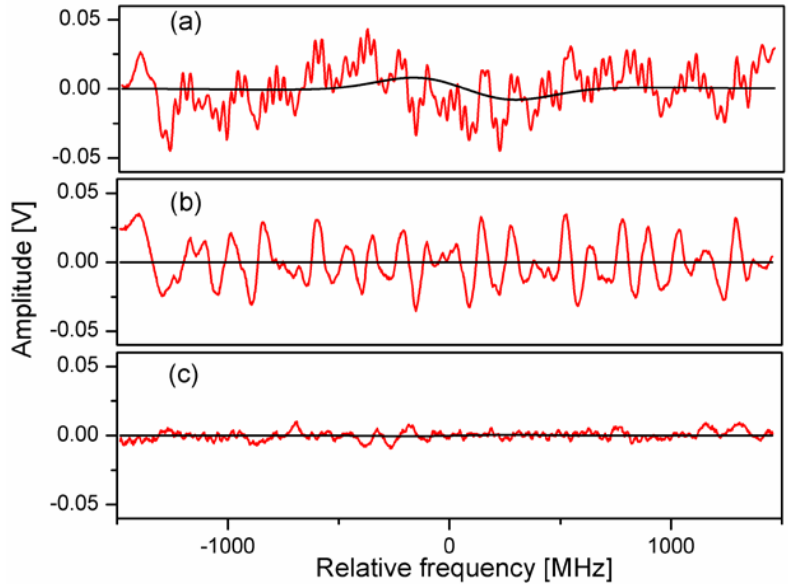

Fig. 2. The drift of the background signal for three modes of detection: (a) detection in transmission without dithering [case *i)*]; (b) balanced detection without dithering [case *ii)*]; and (c) balanced detection with dithering [case *iii)*].

A comparison between the panels (a) and (b) shows, in agreement with the results of Ref. [55], that the balanced detection can remove the wide structure of the background that mainly originates from the EOM and deteriorates the long-term performance of Db NICE-OHMS. It shows that it also eliminates the noise from the narrowest etalons (those with smallest FSR), presumed to originate from interference between the output surface of the output fiber of the laser and the *f*-C. However, as can be seen from panel (b), there is still some remaining background structure, which, as was discussed above, is mainly caused by etalons between one of the cavity mirrors and another surface in the system. In this case, the FSR of the dominant etalon was around 120 MHz, consistent with the distance between the front cavity mirror and the transmission detector. Although such narrow etalon does not significantly impair the Db NICE-OHMS fitting {as can be seen from the black curve in panel (b) and as was discussed in Ref. [55]}, it will distort the line shape of the sD signal and be picked up by the fits to the sD signal. This phenomenon is consistent with the data in Fig. 4 in Ref [49], which displays a simulation of the influence of etalons with different FSRs on the NICE-OHMS signal and show that the analytical signal is highly susceptible to etalons whose FSR is similar to the spectral range of the signal. This does not only indicate that a careful design of the optical system (in this case, with respect to the relative distances between the various components) can reduce the influence of etalons, but also that such a strategy cannot, because of the finite dimensions of the optical components and the length of the system, eliminate all etalons in a system.

On the other hand, as was discussed above, the influence of an etalon can be reduced by dithering either a single component in the system, *viz.* one of the components between which the etalon appears [61], or, as is done in this work, by dithering both mirrors of the cavity and thus the positon of the entire cavity. By dithering the mirrors so that any given etalon was shifted back and forth more than the FSR of the dominant remaining etalon, *viz.* 200 MHz, followed by averaging, the remaining etalons could be significantly reduced. The efficiency of the dithering can be estimated by a comparison between the red curves in the panels (b) and (c) in Fig. 2. As compared to pure transmission NICE-OHMS [panel (a)], the standard deviation of the etalon noise for the novel dithering methodology [panel (c)] was reduced by more than 80 %. This implies that a comparable improvement of the performance of the sD mode of detection is expected.

The detection sensitivities of the different NICE-OHMS strategies for sD detection utilized in this work were quantitatively evaluated by use of the Allan deviation of the absorption coefficient retrieved from fits of a model function of the analytical signal to background signals.

The system was calibrated by the use of 10 ppm $C_2H_2$ at a total pressure of 300 mTorr (for which the sD signal is maximum under the prevalent conditions: the transition addressed and the pertinent intra cavity power). The intra-cavity power was around 40 W and gave rise to a degree of saturation of 16. Figure 3 shows, by the solid curves in blue, black, and red, Allan-Werle plots for the three aforementioned modes of detection investigated in this work at dispersion phase over a time period of 10 min. The dashed line, which has a $\tau^{-1/2}$ dependence (where $\tau$ is the integration time), indicates a white noise level of $1.5 \times 10^{-12}$ cm$^{-1}$ Hz$^{-1/2}$, which corresponds well to the estimated white noise level for the detection schemes *ii)* and *iii)*. This white noise is 4 times above the shot noise level, which is estimated to $3.6 \times 10^{-13}$ cm$^{-1}$ Hz$^{-1/2}$.

The data show that while the background sD NICE-OHMS signal detected in transmission (the blue curve) is affected by drifts for all integration times in the plot (the Allan deviation increases with integration time already from 1 s), the balanced detection (the black curve) can reduce the drifts significantly (the Allan deviation increases with integration time only from around 30 s). As is indicated by the lowermost curve, when also the position of the cavity is dithered, the influence of drift on the system is additionally reduced; the noise in the assessment is in this case significantly affected by drifts for integration times longer than 60 s. For an integration time of 100 s, the amount of drift was reduced with respect to the two other modes of detection [*i*) and *ii*)] by a factor of 20 and 2, respectively. For integration times in the 30 – 70 s range, the system displayed a detection sensitivity of $2.2 \times 10^{-13}$ cm$^{-1}$. For the particular transition addressed, this corresponds to an acetylene concentration of 43 ppt.

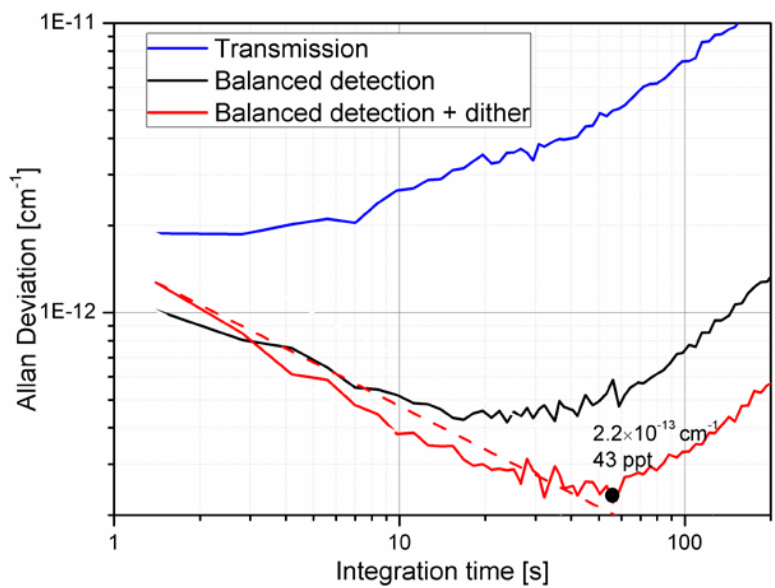

Fig. 3. The Allan plot of the absorption coefficient for sD NICE-OHMS for the case with: (a) detection in transmission without dithering [case *i)*]; (b) balanced detection without dithering [case *ii)*]; and (c) balanced detection with dithering [case *iii)*]. The dashed line represents the estimated white-noise contribution to the case with balanced detection with dithering [case *iii)*].

The slightly higher amount of noise for the shortest integration time (i.e. for 1.4 s) for the case with dithered mirrors in comparison to the case with no dithering can be attributed to periodic noise resulting from the dithering process.

For trace gas measurements the low drift level has some important practical implications. Whenever the highest sensitivity is needed (a detection sensitivity corresponding to around a tenth of a ppb), the signal can be integrated for several tens of seconds before a measurement with an empty cavity (to assess the zero level) needs to be performed (over the same timescale). For the cases when it suffices with a lower sensitivity (in the ppb-range), a multitude of measurements can be performed (in this case for several minutes) before a background calibration needs to be done.

It should be noted that these conclusions could not have been drawn if the system performance would have been assessed only in terms of an ordinary standard deviation since such an analysis does not reflect the influence of drifts of background signals appropriately.

## 4. Application to assessment of acetylene in breath of smokers

As was alluded to above, in order to assess acetylene concentrations in breath down to low- or even sub-ppb levels, high sensitivity LAS techniques such as CRDS [64, 66] and OA-ICOS [67] have previously been employed. The lowest (3σ) detection limit achieved so far is 175 ppt, obtained for a long measurement time, *viz.* 10 min [66]. Since these techniques address Db transitions, they have to rely on finding a $C_2H_2$ transition with as little spectral interference as possible, make the measurements under low pressure (to reduce the effect of pressure broadening), and utilize multi-spectral fitting to extract the concentration. A viable alternative is to use sD detection employing NICE-OHMS, which provides similar detection sensitivity with less stringent requirements.

### *4.1 Simulations*

It was found by simulations based on the HITRAN data base [22], that all $C_2H_2$ transitions that can be addressed by the EDFL suffer from spectral interference by at least one of the afore-mentioned concomitant species. A $C_2H_2$ transition at 6518.4858 cm$^{-1}$ that is markedly affected by a $CO_2$ transition 1.4 GHz away was chosen as the targeted transition in this work. Although the $CO_2$ transition is more than 3 orders of magnitude weaker than the $C_2H_2$ line [having line strengths of $1.8 \times 10^{-24}$ and $8.57 \times 10^{-21}$ cm$^{-1}$/(mol cm$^{-2}$), respectively], the high $CO_2$ concentration in human breath (4 - 6%, which is more than six orders of magnitude higher than the expected $C_2H_2$ concentration in breath) makes the assessment of $C_2H_2$ in breath far from trivial.

Figure 4 shows some spectral simulations of the NICE-OHMS signal from human breath at a pressure of 300 mTorr in the vicinity of the targeted acetylene and $CO_2$ transitions. The red curves display the spectrum from a possible mixture of $C_2H_2$ and $CO_2$ in breath of a smoker (10 ppb $C_2H_2$ and 4% $CO_2$), while the black dots represent the case with no $C_2H_2$, (solely 4% $CO_2$).

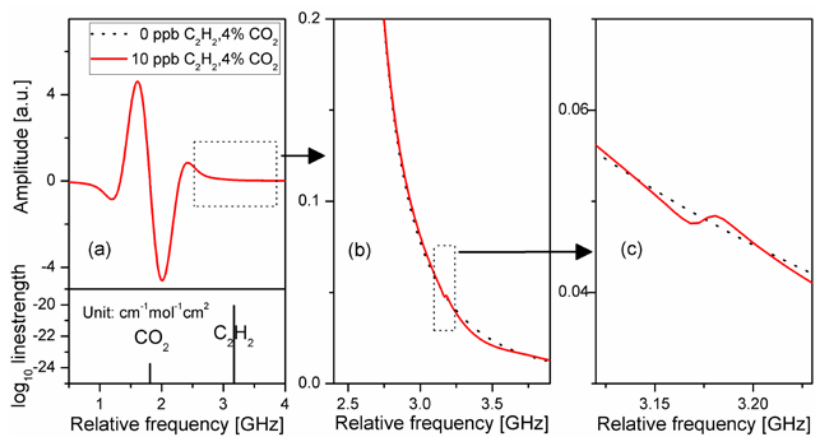

Fig. 4. Red curves: Simulations of expected typical NICE-OHMS signals from smokers' breath, with 10 ppb $C_2H_2$ and 4% $CO_2$ in the vicinity of the $C_2H_2$ transition addressed. Dotted curves: The expected response in the absence of $C_2H_2$.

The upper window of panel (a) displays spectra for Db NICE-OHMS encompassing both the $C_2H_2$ and $CO_2$ transitions for both a typical smoker (the solid red curve) and the corresponding situation in the absence of $C_2H_2$ (dotted black curve, fully overlapping with the solid curve), while the lower shows the positions and the line strengths of the two transitions. Because of the strong absorption of $CO_2$, the $C_2H_2$ signal is drown in the wings of the $CO_2$ to such an extent that it is not possible, on this scale, to distinguish the two curves from each other.

Panel (b) shows a zoom-in of a part of the NICE-OHMS spectrum in the vicinity of the $C_2H_2$ transition. Although both curves are dominated by the Db $CO_2$ response, it is here, in this noise-free simulation of breath, possible to recognize a difference between the two curves. This difference originates from both the expected Db and the expected sD response of $C_2H_2$ (the former represented by the difference in the two NICE-OHMS signals between 2.8 and 3.7 GHz while the latter is given by the narrow feature at around 3.2 GHz, within the dashed box). This shows that to extract information about $C_2H_2$ by Db detection, exceptionally accurate knowledge about $CO_2$ and its transition (not only its concentration and the line strength of the transition, but also broadening parameters) is required. In addition, it can also be concluded that any small amount of noise (in the background or from the signal) will adversely affect the ability to assess a $C_2H_2$ concentration from a Db spectrum with adequate precision. This indicates that it is not advisable to utilize the targeted transition for assessment of $C_2H_2$ in breath by any Db detecting technique.

On the other hand, panel (c) shows a further zoom-in around the center of the $C_2H_2$ line, highlighting the sD response. This shows that the sD response more easily than the Db response can be distinguished from the $CO_2$ signal.

*4.2 Experimental assessments*

The two panels in Fig. 5 show, by the red curves, the sD NICE-OHMS signal of exhaled $C_2H_2$ from two healthy habitual smokers measured 30 minutes after cigarette smoking. End-tidal breath samples (only air from the alveolar region, dead space discarded) were collected in aluminum-coated sampling bags, previously shown to be well-suited for acetylene studies [66]. The bags were then connected to the gas system and the optical cavity was filled with a part of the sample. At the time of analysis, to protect the cavity mirrors, the samples were dehumidified by vacuum-extraction to the cavity via a Nafion tube (PermaPure, PD-100-24MSS), which in a previous work has shown not affect the $C_2H_2$ concentration [66]. The pressure in the cavity was adjusted to 300 mTorr which was found the optimum to maximize the sD signal for the current experimental conditions.

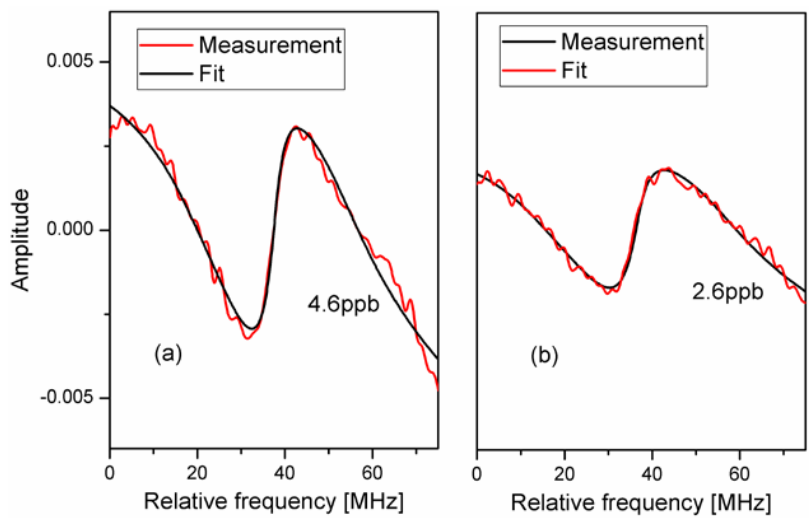

Fig. 5. Sub-Doppler NICE-OHMS signals of $C_2H_2$ in smoker's breath (red) with the corresponding fits (black) under pressure of 300 mTorr. The two panels represent the two individuals.

In order to assess the $C_2H_2$ concentration from the measured signals a fitting procedure was used to separate the sD $C_2H_2$ response from the Db responses of $CO_2$ and $C_2H_2$. The fitting of the sD response was based on an empirical line shape model retrieved by sD NICE-OHMS from a standard reference gas (10 ppm of $C_2H_2$ in $N_2$) measured under similar conditions as the breath analyses were performed, i.e. with the same intracavity power and at the same pressure, 300 mTorr. It was assumed that the pressure broadenings from the ambient air and breath are similar to that of the standard reference gas (i.e. $N_2$). Because of this, the fitting to the measured data could be done with only a limited number of parameters; the sD $C_2H_2$ response could be fitted with only a single parameter (the $C_2H_2$ concentration), while, since the data considered for the analysis is only covering a small part of a Db feature (a part of its wing), the fits to the Db response of $CO_2$ were taken as phenomenological second-order polynomials.

As can be seen from Fig. 5, the combined fits (of the sD and the Db signals, represented by the black curves) pick up virtually all of the sD $C_2H_2$ signals. This good agreement justifies the assumptions made for the empirical line shape model for the Db signal. Based on the fitting routine, the concentrations of $C_2H_2$ in the breaths of the two healthy smokers could be assessed to 4.6 and 2.6 ppb, respectively, which are in good agreement with the results reported in previous breath $C_2H_2$ studies for samples analyzed 30 minutes after smoking [66, 67].

An assessment of the strength of the remaining Db response resulted in a $CO_2$ concentration of 5.3%, which is consistent with a normal concentration of $CO_2$ in human breath. A further analysis of the detection sensitivity of $CO_2$ detected by the means used in this work was assessed by the use of an Allan plot (not shown). It was found that the minimum detectable $CO_2$ concentration by this means of detection was 2 ppm, significantly below the assessed concentration. This indicated that also the assessment of $CO_2$ was adequate.

It is worth to note that we neither found any detectable amount of $C_2H_2$ in the breath of non-smokers, nor did we find it in the outdoor air. This is in contrast to the findings of Ref. [66], performed in Helsinki, Finland, where background levels of around 1 ppb were found in both situations. We attribute these differences to cleaner air in Umeå than in Helsinki; a lower population and a lower degree of industrial activities (transportation and air pollution). The result also confirms that acetylene is not endogenously produced in healthy human subjects.

## 5. Summary and conclusions

In summary, a sensitive spectrometer for trace gas detection with high spectral resolution based on cavity-position-modulated sub-Doppler (sD) NICE-OHMS has been developed. For best performance, the instrument was based on a system for balanced detection NICE-OHMS. Although this system previously has demonstrated shot-noise-limited conditions for Doppler-broadened (Db) detection [55], it did not provide the corresponding conditions for sD detection. The reason was attributed to etalons generated between one of the cavity mirrors and a second surface in the system with a periodicity similar to the width of the sD NICE-OHMS signal.

To improve on the performance of the system for sD detection, the influence of the remaining etalons was reduced by providing two sinusoidal signals (90 Hz) to the two PZT crystals onto which the cavity mirrors were glued so as to dither the position of the cavity without affecting its physical length. This implies that all etalons that appeared between one of the mirrors of the cavity and any other surface in the system were actively dithered. By then averaging the signal over time, the influence of the remaining etalons could be significantly reduced (by 80 % with respect to the signal detected in ordinary transmission NICE-OHMS).

According to an Allan deviation analysis, the system can provide a sD detection sensitivity of $2.2 \times 10^{-13}$ cm$^{-1}$ for an integration time of 60 s, which is 20 times better than when the signal is detected in transmission (by use of conventional NICE-OHMS) and 2 times better than when only balanced detection was used. For the $C_2H_2$ transition addressed, this corresponds to a $3\sigma$ detection limit of 130 ppt, which marginally supersedes the so far lowest ($3\sigma$) detection limit obtained for a cavity enhanced based LAS technique (175 ppt, assessed for a long measurement time, *viz.* 10 min ) [66].

It was then shown that, by the use of sD detection, the assessment was not significantly affected by spectral interference from a $CO_2$ transition positioned 1.4 GHz from the targeted $C_2H_2$ transition despite the fact that the $CO_2$ concentration was more than six orders of magnitude higher than that of $C_2H_2$. For ordinary Db detection, the $C_2H_2$ transition is practically not detectable in the presence of $CO_2$. However, when the instrumentation was run for detection of sD features, it was clearly visible and quantifiable.

As a demonstration of the applicability of the instrument, the acetylene concentrations in the breath of two healthy smokers, measured half an hour after cigarette smoking, were assessed. The $C_2H_2$ concentrations in the smokers' breath were assessed to 4.6 and 2.6 ppb, respectively. Based on the Db response, a procedure was worked out for simultaneous assessment of $CO_2$. We did neither find any detectable amount of $C_2H_2$ in the breath of non-smokers, nor did we find it in the outdoor air.

Based on the demonstration given, we predict that sD NICE-OHMS is more resilient to spectral interferences than cavity-enhanced techniques addressing Db spectra, e.g. cavity ring down spectroscopy and off-axis integrated cavity output spectroscopy. Thus, sD NICE-OHMS has great potential in applications where spectral interference is a limiting factor and ultra-high sensitivity is required, as is the case in breath gas analysis.

**Funding**

This project was supported by the Swedish Research Council (Vetenskapsrådet, VR: OA: 2015-04374; FMS: 2013-6031), the Kempe foundations (OA: JCK-1317; FMS: SMK-1446) and the National Natural Science Foundation of China (61675122).

**Acknowledgements**

We acknowledge Umeå University's program "Strong research environments" for support. GZ thanks the China Scholarship Council (CSC) for support.